Accepted at *IEEE* Ubiquitous Computing, Electronics, and Mobile Communication (UEMCON), 2026

# Synthetic Fingerprints for Children Under Four: Generation and Biometric Evaluation

Faheem Ahmad*, Ajan Ahmed*, Mst Rumana Sumi†, Stephanie Schuckers§, and Masudul Imtiaz‡
*†‡Electrical and Computer Engineering, Clarkson University, Potsdam, NY, USA
§Computer Science, University of North Carolina at Charlotte, Charlotte, NC, USA
Email: {*fahmad, *aahmed, †sumima, ‡mimtiaz}@clarkson.edu, §sschucke@charlotte.edu

***Abstract*—Fingerprint recognition in children under four is of interest for longitudinal identity applications, but research in this age range is constrained by the limited availability and sensitivity of real fingerprint data. Synthetic data may provide a useful complementary resource if generated samples are carefully evaluated for biometric quality, similarity to the real training data, and identity diversity. This paper presents an evaluation and selection protocol for synthetic fingerprints generated from a small pediatric dataset. The protocol was applied to 32,000 candidates produced by an age-conditioned generator fine-tuned with 205 fingerprints from nine children. Candidates were evaluated using NFIQ 2, NBIS minutiae extraction and matching, fingerprint-pattern classification, similarity to the complete real reference set, and pairwise similarity among retained synthetic samples. After candidate filtering and a final symmetric pairwise verification, 1,985 synthetic fingerprints remained. The youngest age group continued to produce retained samples within the sampling budget, whereas the oldest original age group produced 24 retained prints. A data-derived two-group age representation improved pattern-distribution agreement for the younger group. Repeated renderings of fixed synthetic identities produced minutiae-based mated scores comparable to the real mated scores, although a DINOv2 texture representation showed substantially greater within-generator similarity. The results indicate that synthetic pediatric fingerprints can support controlled research use, but their evaluation should include both real-to-synthetic similarity and synthetic-to-synthetic diversity, and conclusions should remain specific to the matchers and measurements used.**

## I. Introduction

Fingerprint recognition in early childhood has been investigated for applications such as vaccination tracking, birth registration, and family reunification, particularly when conventional identity documents are unavailable or difficult to maintain [1]–[3]. At the same time, research involving fingerprints from very young children is challenging. Longitudinal collections require repeated participation, and the resulting biometric data are sensitive and are rarely available for broad research use [4], [5]. Recognition is also affected by rapid changes in finger size, ridge spacing, usable contact area, skin deformation, and image quality during the first years of life [6], [7].

This work was supported by the Center for Identification Technology Research (CITeR) and under NSF Grants 2413228 and 2501916.

Synthetic fingerprint generation offers one possible way to supplement these limited datasets. Adult fingerprint synthesis has progressed from explicit ridge-field models such as SFinGe [8] to learned methods such as PrintsGAN [9] and diffusion-based GenPrint [10]. These methods are evaluated primarily for realism and matching utility on large datasets. For pediatric data, however, an additional concern arises: a model trained on a small number of children may reproduce features that are unusually similar to its training examples, and generated samples may also contain repeated or near-duplicate synthetic identities. In this study, the age condition is a discrete age-group label supplied to the generator as its class input during fine-tuning and sampling, so that samples can be requested for a specific age group. This label directs generation toward images associated with that group, but it does not by itself necessarily imply that the resulting ridge structure follows the same age-related trends observed in the real data.

This study therefore emphasizes evaluation of the generated collection rather than proposing a new generator architecture. An existing conditional fingerprint generator is fine-tuned on fingerprints from children under four and is used to produce candidate samples. Each candidate is then assessed for image quality, minutiae extraction, similarity to the real reference set, fingerprint-pattern structure, and similarity to previously retained synthetic fingerprints. A final pairwise verification is performed after sampling because the NBIS matcher used in this study is not fully symmetric with respect to input order. The study addresses three questions. First, how many distinct synthetic samples can be retained when similarity to the real data and similarity among synthetic samples are considered simultaneously? Second, do calendar-based or data-derived age groups better reflect the observed fingerprint structure in this small dataset? Third, do multiple renderings of a fixed synthetic identity produce comparison-score distributions that resemble repeated impressions of a real finger? The main contributions are: (1) a reproducible evaluation and selection protocol for a small pediatric fingerprint generator; (2) an analysis of sample yield and synthetic identity diversity over 32,000 generated candidates; (3) a comparison of calendar-based and data-derived age conditions; and (4) a repeated-impression analysis using both a minutiae matcher and a texture embedding. The final collection contains 1,985 synthetic fingerprints and is intended for controlled research use together

TABLE I
REPRESENTATIVE SYNTHETIC FINGERPRINT GENERATORS AND EVALUATION SCOPE.

| Generator | Child data | Age control | Real-reference test | Synthetic duplicates |
|---|---|---|---|---|
| SFinGe [8] | no | no | no | no |
| CFG [11] | no | no | vs. chance | no |
| CFG V2 [13] | under 9 | coarse groups | not reported | no |
| PrintsGAN [9] | no | no | vs. chance | at scale [10] |
| GenPrint [10] | no | no | vs. chance | at scale |
| **This work** | 0.59–4.23 yr | data-derived groups | vs. control rate | pairwise filtering |

with the accompanying provenance and evaluation records.

## II. Related Work

### A. Fingerprints of Young Children

Studies of infant and young-child fingerprints have demonstrated that reliable recognition is possible under appropriate acquisition conditions, while also showing that early-childhood fingerprints cannot simply be treated as scaled adult fingerprints. Prior work includes recognition studies above approximately six months [1], national-registry experiments beginning at birth [2], dedicated infant-capture hardware [3], longitudinal evaluations [4], and analyses of age-dependent image quality [6]. These studies motivate methods that can reduce reliance on unrestricted access to real pediatric biometric data while preserving useful biometric characteristics for research.

### B. Synthetic Fingerprints and Their Evaluation

Synthetic fingerprint generation has evolved from handcrafted models [8] to conditional generative approaches [9], [11] and controllable diffusion models [10]. Most work has focused on adults. Age-aware synthetic biometrics have been investigated more extensively for face recognition [12]. For fingerprints, CFG V2 includes an under-nine category [13], CFG V2 extends the original Clarkson Fingerprint Generator into a conditional GAN whose class labels cover finger type and three age groups (under nine, nine to fourteen, and adult), making under nine the finest age granularity available in existing fingerprint generators. But this grouping is substantially broader than the rapid developmental interval considered here. The generator used in our study is a fine-tuned version of the conditional model of Abbas *et al.* [14], with adaptive discriminator augmentation (ADA) [15].

Evaluation of synthetic biometric data commonly considers visual fidelity, distributional similarity, similarity to training data, and duplicate identities [10], [11], [16]. The present work combines these ideas with standard fingerprint instruments: NFIQ 2 for image quality [17], NBIS for minutiae extraction and matching [18], Poincaré-index singularity analysis for pattern classification [19], [20], the standard fingerprint pattern taxonomy [21], and DINOv2 as a secondary texture representation [22]. Table I summarizes the emphasis of this study relative to representative synthetic fingerprint generators.

## III. Materials and Methods

### A. Pediatric Fingerprint Dataset

The experimental design distinguishes between the persistence of the underlying ridge pattern and the appearance of an acquired fingerprint sample. Ridge patterns are expected to remain stable after formation [23], but an image collected from a very young child may contain a smaller usable area, fewer extracted minutiae, lower contrast, and stronger deformation than an image from an older child. In the present dataset, the median minutiae count increases from 43 in the 0–1.5-year group to 96 in the 3-years-and-older group. Median print area increases from 131 to 239 $mm^2$, ridge period from 0.38 to 0.43 mm, and NFIQ 2 quality from 12 to 33.5. These observations support age-aware evaluation of the generated samples.

The dataset contains 205 preprocessed fingerprint samples from nine children collected between 0.59 and 4.23 years of age employing a custom fingerprint scanner. High-resolution samples were acquired across multiple visits, resampled to 500 pixels per inch, and placed on a 512-pixel canvas. These fingerprints are used for generator fine-tuning, threshold estimation, similarity evaluation, and statistical comparison.

### B. Generator and Candidate Generation

Because 205 samples are insufficient for training a high-capacity generator from scratch, an existing conditional StyleGAN2 fingerprint generator was used for initialization [14]. The base model had been trained on 20,844 prints from 338 individuals. It was fine-tuned using ADA to reduce overfitting during transfer to the much smaller pediatric dataset [15]. During transfer, 152 of 153 generator tensors were retained and the class embedding was relearned for 26 classes defined jointly by finger and age group.

A base-generator control was used to characterize background real-reference matching. Prints from the base generator, which had not been trained on the nine children, were compared with the same real gallery using identical preprocessing and matching. This provided a practical reference rate for interpreting real-to-synthetic similarity. Candidate samples were then drawn from the 20-, 40-, and 60-kimg checkpoints and from the data-derived age-group model described below. The 200-kimg checkpoint was retained for comparison because it provided the best aggregate distributional fidelity, although its real-reference matching rate was higher.

### C. Fingerprint Comparison Protocol

Mated comparisons contain two impressions of the same finger, whereas non-mated comparisons contain different fingers [24]. Minutiae were extracted with NBIS `mindtct`, and similarity scores were produced with `bozorth3` [18]. Finger-image quality was measured with NFIQ 2 [17], [25]. For similarity score $s(x, y)$, larger values indicate greater similarity. The primary operating threshold was

$$\tau_{0.999} = \min\{t \in \mathbb{Z} : t \geq Q_{0.999}(S_{NM})\} = 26, \qquad (1)$$

where $S_{NM}$ contains 18,127 real non-mated scores. The observed false-match rate at this threshold was 0.14%. Because this estimate is obtained from the tail of a small pediatric dataset, the conventional `bozorth3` threshold of 40 is also reported as a stricter supplementary operating point. The base-generator control produced a 4.50% real-reference match rate at threshold 26 and 0 of 1,001 matches at threshold 40.

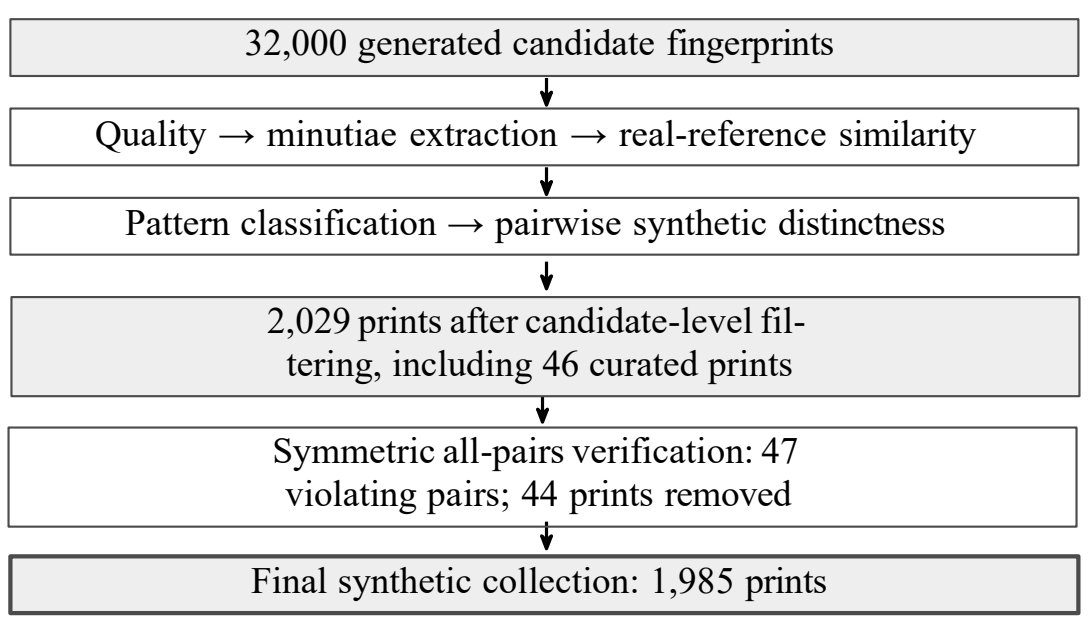

Fig. 1. Candidate evaluation and selection workflow. Criteria are applied sequentially, followed by a symmetric pairwise verification of the retained collection.

### *D. Candidate Evaluation and Selection*

Let $x_i$ denote a generated candidate and $A_{i-1}$ the set of synthetic prints retained before evaluating $x_i$. A candidate is retained when

$$C_q(x_i) \wedge C_m(x_i) \wedge C_r(x_i) \wedge C_p(x_i) \wedge C_d(x_i, A_{i-1}) = 1, \quad (2)$$

where the five criteria correspond to image quality, successful minutiae extraction, real-reference similarity, pattern classification, and pairwise synthetic distinctness. The procedure is summarized in Fig. 1.

First, the candidate must meet or exceed the median NFIQ 2 score of the real fingerprints in the corresponding age group. Second, NBIS must extract minutiae. Third, the candidate must remain below thresholds 26 and 40 against each of the 205 real fingerprints. Fourth, a fingerprint pattern must be assigned using a Poincaré-index rule set [19]–[21]. Fifth, the candidate must score below 15 against every previously retained synthetic print. Since `bozorth3` may produce different scores when the input order is reversed, the complete retained set is subsequently checked in both score directions for every pair. Image quality, minutiae extraction, and pattern classification are evaluated before the more expensive gallery and pairwise comparisons. The real-reference comparison identifies candidates that are unusually similar to the available real samples under NBIS, while the pairwise comparison limits repeated or near-duplicate synthetic identities under the same matcher. Neither comparison establishes a matcher-independent property; the resulting collection is therefore characterized with respect to the stated preprocessing, scores, and thresholds.

### *E. Data-Derived Age Groups*

The original conditioning used calendar groups of 0–1.5, 1.5–3, and 3–5 years. A data-derived alternative was also evaluated because the available real fingerprints did not provide strong empirical support for the 1.5-year boundary. K-means clustering of print-level ages was evaluated with the silhouette criterion [26], subject to two practical constraints: each group had to contain at least two children and at least 15 images. Under these constraints, a two-group split near age 2 was selected. Leave-one-child-out analysis was used to evaluate stability. In all nine folds, two groups were selected and no retained sample changed group, giving an adjusted Rand index of 1.0 [27]. The boundary midpoint was 2.075 years except when a child near the gap was omitted.

The original and data-derived conditions were compared using two matched training configurations. Trial A used the original calendar labels but was evaluated in the two-group representation. Trial B was fine-tuned directly with the data-derived two-group labels, producing 18 classes, while the remaining training configuration was unchanged. This comparison is intended to assess which representation better matches the observed structure of this dataset.

### *F. Sampling Yield and Repeated-Impression Evaluation*

The retained set was initialized with 46 previously curated synthetic prints. Each additional candidate was generated from a seeded latent input not used in earlier probes, and each newly retained print was included in the pairwise comparison set for later candidates. Sampling for a group was stopped after 600 consecutive rejections. A total of 32,000 candidates was evaluated. The initial pilot used 10,640 candidates from the 60-kimg checkpoint: 5,040, 4,800, and 800 from the youngest, middle, and oldest original age groups. The expansion included 21,360 additional candidates: 9,960 for the youngest group, 3,900 each from the 20- and 40-kimg checkpoints, and 3,600 from the data-derived model.

A separate experiment examined repeated renderings of fixed synthetic identities. For 82 fresh identities that satisfied the same criteria, six impressions were generated: one canonical rendering, two renderings with resampled generator noise, and three with simulated skin distortion. The distortion model included rotation, translation, elastic warp, and contrast change. Distortion strength was selected to minimize Wasserstein distance between canonical-to-distorted scores and the real mated-score distribution. One distorted rendering failed minutiae extraction, leaving 491 impressions. Pairs formed from two renderings of the same synthetic identity were treated as synthetic mated pairs, and pairs formed from different synthetic identities were treated as synthetic non-mated pairs, following the terminology of ISO/IEC 19795-1 [24]. Both categories were evaluated with NBIS and a DINOv2 texture embedding [22]. NBIS measures consistency of minutiae-based identity information, whereas DINOv2 provides a complementary image-level comparison. Every generated impression was also compared with the complete real reference set.

## IV. Results

### *A. Why Multiple Evaluation Criteria Are Needed*

The 60-kimg and 200-kimg checkpoints illustrate why distributional fidelity, real-reference similarity, and synthetic diversity should be examined separately. For the 60-kimg checkpoint, within-group synthetic pairs match each other at rates of 3.1 to 15.2% at threshold 26. The 200-kimg checkpoint gives 3.1 to 6.5%. Both are above the corresponding same-group real non-mated rate of 0.28%. At threshold 40, no real

TABLE II

PER-PRINT STATISTICS (MEAN ± STANDARD DEVIATION) OF REAL AND FINAL SYNTHETIC FINGERPRINTS BY ORIGINAL AGE GROUP; ENTRIES IN PARENTHESES GIVE REAL / SYNTHETIC SAMPLE COUNTS. THE 218 PRINTS FROM THE DATA-DERIVED TWO-GROUP MODEL ARE NOT INCLUDED BECAUSE THE GROUPING DIFFERS. REAL NFIQ 2 VALUES ARE AVAILABLE FOR 35, 50, AND 114 PRINTS; SIX PRINTS HAD INSUFFICIENT USABLE RIDGE AREA.

| Age group | Metric | Real | Synthetic |
|---|---|---|---|
| 0–1.5 y (37 / 1,640) | Minutiae count | 47.1 ± 30.2 | 16.7 ± 12.8 |
| | NFIQ 2 quality | 10.8 ± 8.1 | 20.4 ± 7.9 |
| | Print area (mm$^2$) | 130.1 ± 56.4 | 67.1 ± 26.4 |
| 1.5–3 y (51 / 103) | Minutiae count | 63.1 ± 22.5 | 29.7 ± 21.8 |
| | NFIQ 2 quality | 29.4 ± 17.9 | 40.2 ± 12.6 |
| | Print area (mm$^2$) | 180.4 ± 59.1 | 117.3 ± 42.8 |
| 3+ y (117 / 24) | Minutiae count | 94.0 ± 37.4 | 62.9 ± 28.9 |
| | NFIQ 2 quality | 35.5 ± 19.9 | 54.9 ± 15.2 |
| | Print area (mm$^2$) | 245.6 ± 94.6 | 141.6 ± 92.0 |

pair and no base-generator control print matches, while 0.87 to 6.09% of 60-kimg synthetic pairs still match.

The real-reference matching behavior differs. The 60-kimg checkpoint is at or below the base-generator control rate in every group (0.67 to 0.96 times), while the 200-kimg checkpoint reaches 2.0 to 4.0 times the control rate. Mean Kolmogorov–Smirnov distance over six fingerprint statistics is 0.21 for the 60-kimg checkpoint and 0.16 for the 200-kimg checkpoint when pooled. Thus, the checkpoint with better aggregate distributional fidelity does not necessarily provide lower real-reference similarity or greater synthetic identity diversity.

### *B. Properties of the Final Synthetic Collection*

Table II compares the final synthetic collection with real prints from the same original age groups. Since retained candidates were required to meet the median real NFIQ 2 value of their corresponding group, their average quality is higher than that of the real data. This is a direct consequence of the selection criterion.

The synthetic prints also contain fewer minutiae and smaller print areas than the real prints. The difference is largest in the youngest group, where the retained prints contain 17 minutiae on average compared with 47 in the real data. Samples with fewer minutiae are less likely to produce high comparison scores, so the pairwise distinctness criterion favors lower-information samples. This effect is important when interpreting the large number of retained samples in the youngest group. Example synthetic fingerprints are shown in Fig. 2.

### *C. Repeated Impressions of Fixed Synthetic Identities*

Fig. 3 shows six renderings of two fixed synthetic identities. The main ridge structure remains stable, while generator noise modifies texture and the distortion model changes pose and contrast. With the minutiae matcher, the synthetic mated scores overlap the real mated distribution. The synthetic mated median is 7, with quartiles 0 and 17, compared with a real mated median of 9 and quartiles 5 and 24.5. A one-sided Mann–Whitney $U$ test for excess gives $p \approx 1.0$, with

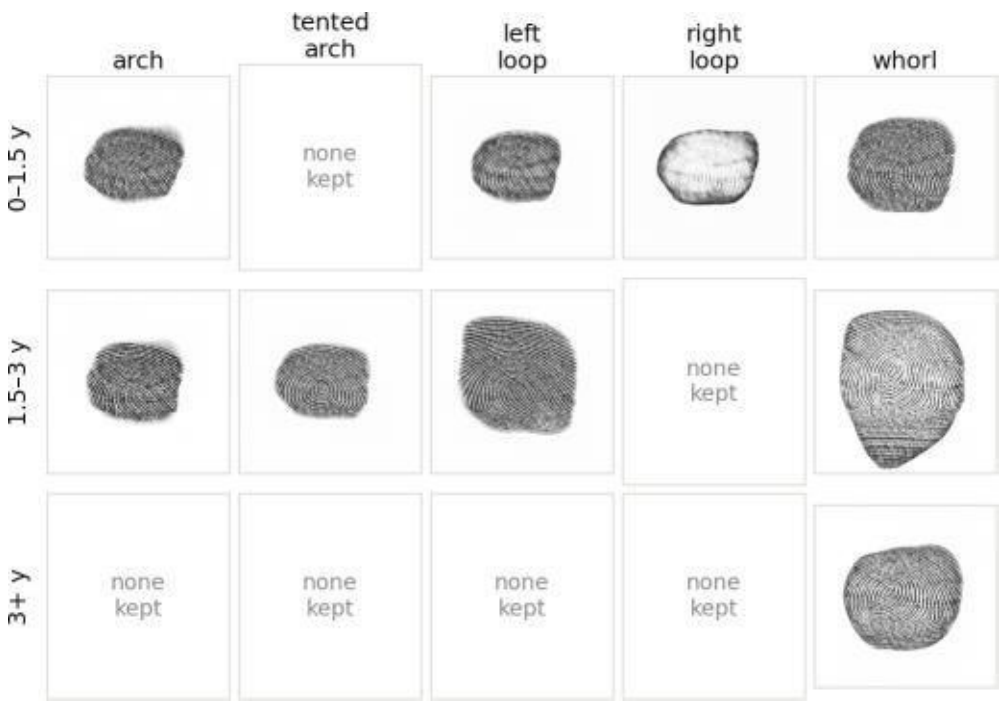

Fig. 2. Examples from the final synthetic collection. Samples are selected from the curated set by NFIQ 2 quality and fingerprint-pattern category.

TABLE III

COMPARISON OF AGE-CONDITIONING SCHEMES IN THE TWO-GROUP REPRESENTATION. $p$ IS THE PERMUTATION-TEST VALUE FOR AGREEMENT WITH THE REAL PATTERN DISTRIBUTION. REAL-REFERENCE VALUES ARE RELATIVE TO THE 4.50% BASE-GENERATOR CONTROL RATE AT THRESHOLD 26; TRIAL A REPORTS THE RANGE ACROSS THE ORIGINAL GROUPS, WHEREAS TRIAL B REPORTS YOUNGER / OLDER VALUES.

| Model | kimg | Younger $p$ | Older $p$ | Real-ref. fold |
|---|---|---|---|---|
| A: calendar | 60 | 0.005 | $< 10^{-4}$ | 0.67–0.96 |
| B: derived | 60 | 0.32 | $< 10^{-4}$ | 1.28 / 0.89 |
| B: derived | 200 | 0.41 | $2.5 \times 10^{-4}$ | 2.24 / 4.23 |

Kolmogorov–Smirnov distance 0.33 and Wasserstein distance 8.4. Synthetic non-mated scores remain below the real non-mated scores, with medians 0 and 7, respectively. No generated impression reaches the operating match threshold against any real fingerprint with this matcher.

The texture-based analysis gives a different result. Synthetic mated DINOv2 cosine similarity has median 0.94, above the real mated median of 0.74, and synthetic non-mated similarity has median 0.80. The generated impressions therefore retain a common texture characteristic that is not captured by the minutiae matcher. This limits the current collection for experiments that rely directly on generic image embeddings and motivates fingerprint-specific texture evaluation in future work.

### *D. Comparison of Age-Conditioning Schemes*

Table III compares the original calendar labels with the data-derived age groups. For the younger group, the calendar-conditioned model differs from the real pattern distribution at 60 kimg ($p = 0.005$). The data-derived model is more consistent with the real pattern distribution at 60 kimg ($p = 0.32$) and 200 kimg ($p = 0.41$). For this dataset, the two-group representation therefore provides a better description of the younger-group pattern distribution.

Both configurations show an excess of whorl patterns in the older group ($p \le 2.5\times10^{-4}$). Pattern agreement is strongest at 200 kimg, but the real-reference matching rate is also highest at that checkpoint. The 60-kimg checkpoint was therefore used for the final collection. These results suggest that age conditioning should be checked against measured fingerprint structure rather than relying only on predetermined calendar intervals.

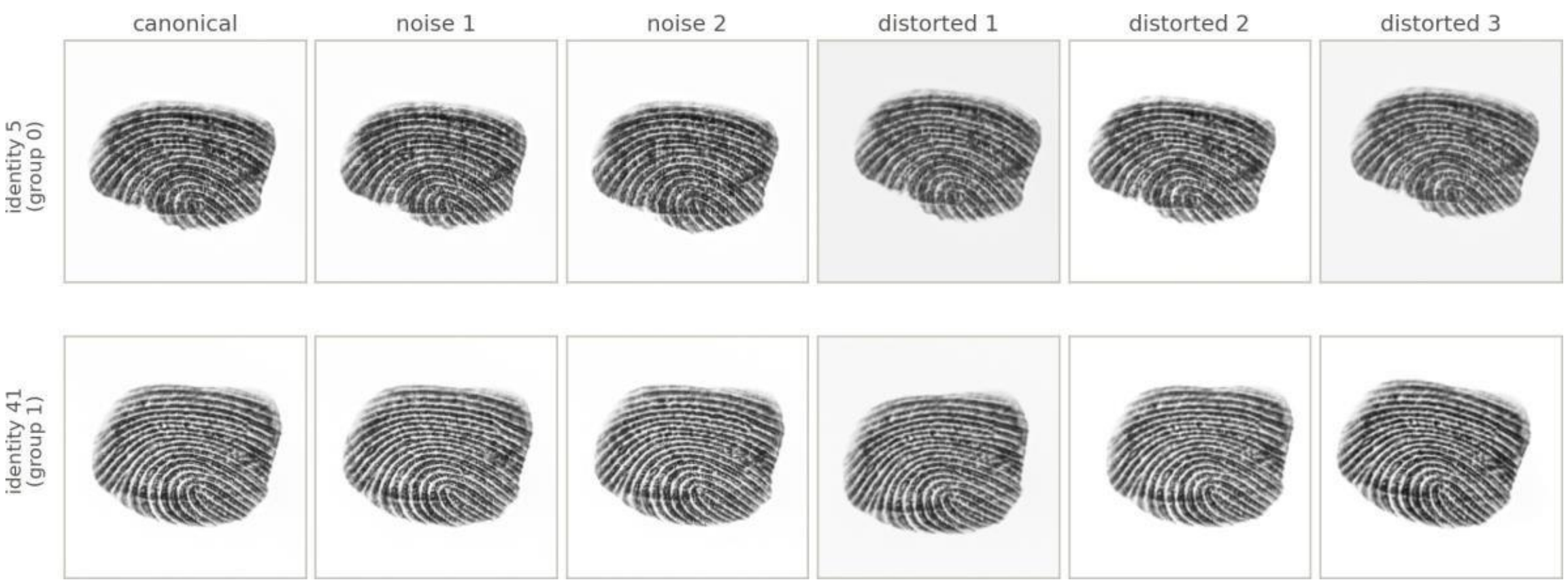

Fig. 3. Six renderings of two fixed synthetic identities: a canonical image, two images with resampled generator noise, and three images produced with the calibrated distortion model.

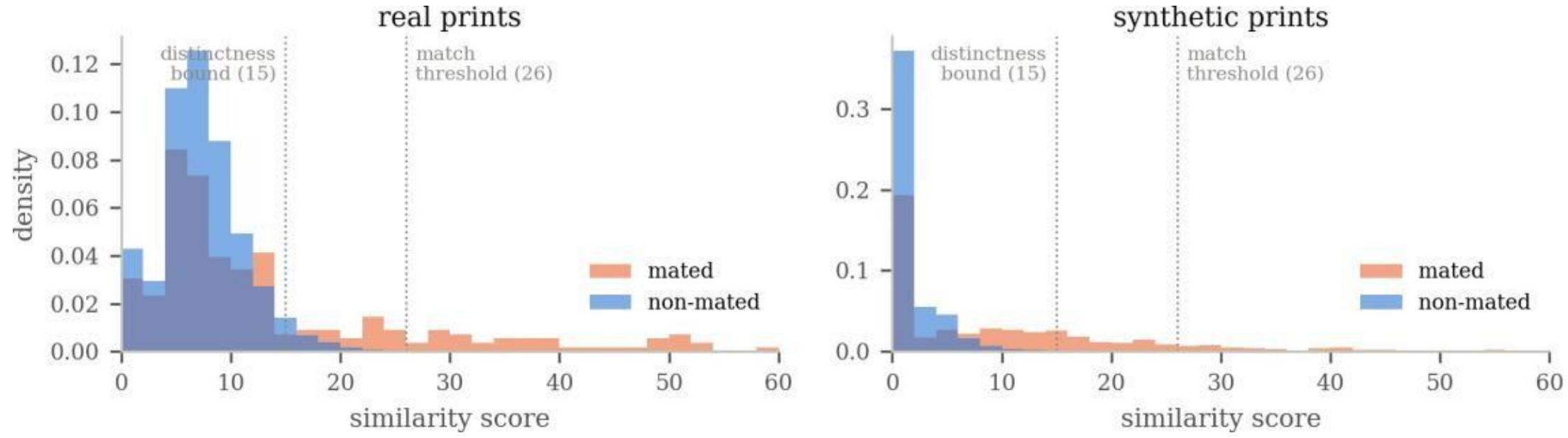

Fig. 4. BOZORTH3 score distributions for real pairs (left) and synthetic pairs (right), with the pairwise distinctness bound (15) and operating threshold (26). The axis ends at 60; 31 real mated and 62 synthetic mated pairs have higher scores, with maxima of 241 and 313.

### *E. Sampling Yield and Final Pairwise Verification*

Fig. 5 summarizes the samples retained from 32,000 candidates. The youngest original age group did not reach the 600-consecutive-rejection stopping criterion. Its acceptance rate remained approximately 8 to 21%, and 935 additional prints were retained during the expansion. Its reported count should therefore be interpreted as the yield obtained within the present sampling budget rather than as a maximum. The middle and oldest groups reached the stopping criterion earlier, with 78 and 4 retained identities from the 60-kimg checkpoint. Additional sampling from the 20- and 40-kimg checkpoints and from the data-derived model increased the oldest original-group count to 24 and added 207 and 17 prints in the two data-derived groups before the final pairwise verification.

Table IV shows the first criterion responsible for removing each candidate. Pairwise synthetic similarity is the largest source of rejection, accounting for 19,544 candidates (61.08%), followed by the quality criterion with 9,675 candidates (30.23%). Similarity to the real reference set removes 587 candidates (1.83%) that had already passed quality and minutiae extraction, showing that checkpoint-level comparisons alone do not identify every high-similarity sample.

The symmetric all-pairs verification was applied to 2,029 retained prints and identified 47 pairs at or above the pair-wise distinctness bound. Removing the later member of each violating pair removed 44 prints and left 1,985. The highest remaining pair score was 14. A reverse-direction comparison, in which each real print was scored against the complete synthetic collection, produced no score at or above the operating threshold of 26; the maximum was 25. These findings characterize the final collection for the specific NBIS matcher, thresholds, and preprocessing used here.

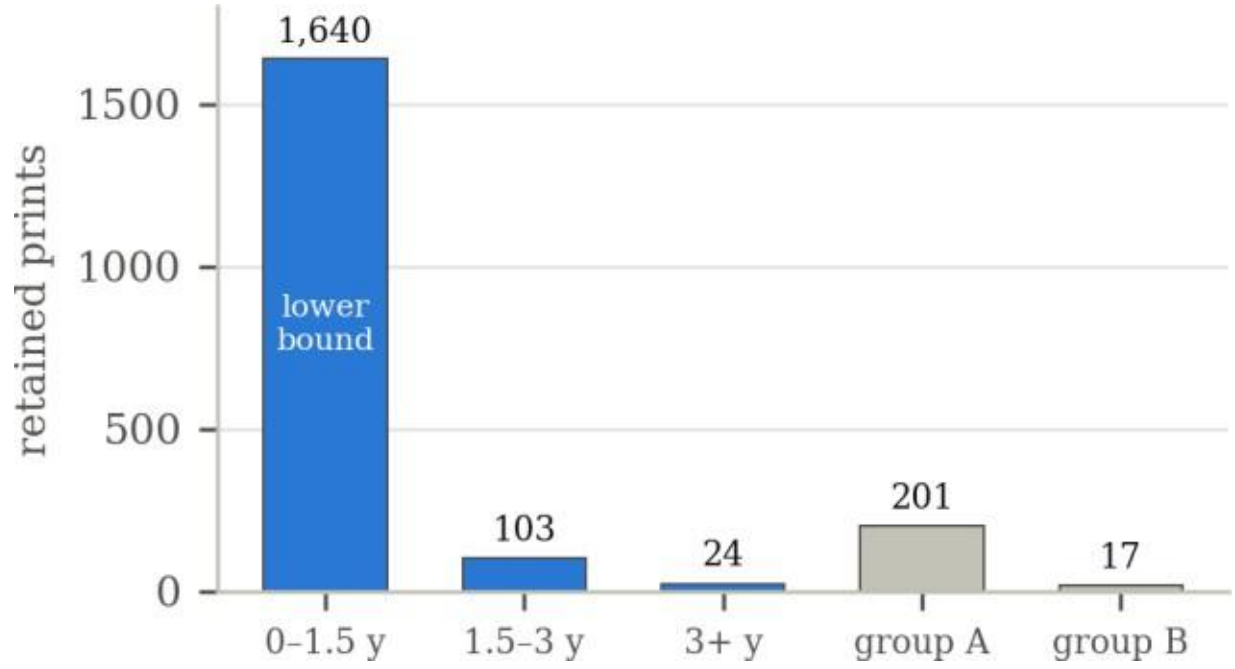

Fig. 5. Final synthetic sample counts by age group (1,985 total). The youngest original group did not reach the sampling stopping criterion; groups A and B are from the data-derived two-group model.

The final collection contains 1,640, 103, and 24 prints in

TABLE IV

FIRST CRITERION RESPONSIBLE FOR REMOVING EACH OF THE 32,000 CANDIDATES. THE 1,983 NEWLY RETAINED CANDIDATES PLUS 46 CURATED PRINTS PRODUCED THE 2,029 PRINTS ENTERING THE FINAL PAIRWISE VERIFICATION.

| Outcome | Candidates | Share |
|---|---|---|
| Quality below real-group median | 9,675 | 30.23% |
| Minutiae extraction failed | 4 | 0.01% |
| High similarity to real reference set | 587 | 1.83% |
| Pattern not classified | 207 | 0.65% |
| High similarity to retained synthetic prints | 19,544 | 61.08% |
| Retained | 1,983 | 6.20% |

the three original age groups and 201 and 17 prints in the two data-derived groups, for a total of 1,985. Each sample is accompanied by its source checkpoint, NFIQ 2 score, minutiae count, geometry, pattern label, and similarity measurements, and the complete 32,000-candidate evaluation log is retained with the collection.

## V. DISCUSSION

The results support several practical observations for synthetic biometric data constructed from small pediatric datasets. First, similarity to the real training data and similarity among generated samples are different properties. The 60-kimg model has a real-reference matching rate near the base-generator control, yet the generated candidates still include many mutually similar pairs. Conversely, the 200-kimg model produces better aggregate distributional fidelity but substantially greater real-reference matching. Reporting only one of these measurements would therefore give an incomplete characterization of the generated data.

Second, the selection protocol changes the composition of the synthetic collection. The final samples contain approximately one third to two thirds of the real minutiae count in the corresponding original age groups, and the youngest group is particularly sparse. This explains part of the high sample yield for that group and means that the collection should not be treated as a complete statistical substitute for real pediatric fingerprints. It is better suited to controlled experiments in which quality, geometry, and matcher behavior are explicitly considered. Appropriate uses include method development, comparison of preprocessing strategies, and analysis of how quality measures interact with matcher scores. The collection is not intended to reproduce population frequencies or replace longitudinal pediatric data. Because samples were selected using NFIQ 2 and NBIS, studies that use the same instruments should also recognize the resulting selection bias and, where possible, include an independent matcher or representation.

Third, conclusions about real-to-synthetic similarity and synthetic distinctness are matcher-dependent. The minutiae matcher recognizes only about one third of genuine child-print pairs in this setting. The DINOv2 experiment also shows that samples judged appropriately separated in minutiae space may remain unusually similar in a generic texture representation. Evaluation with additional commercial and open-source matchers, as well as fingerprint-specific embedding models, is therefore an important next step.

The study has additional limitations. Only nine children are represented, so age and identity effects cannot be fully separated, and each generator configuration is based on one training run. The leave-one-child-out analysis demonstrates that the selected two-group partition is not driven by a single participant, but the boundary remains specific to this dataset. The data-derived model also continues to overproduce whorl patterns in the older group. Finally, the checkpoint with the best distributional measures (FID 105.9, KID 0.054) produces real-reference matching at approximately three times the control rate, emphasizing that distribution-level image statistics cannot substitute for biometric comparison. Future work should include multi-matcher evaluation, independent ablation of the selection criteria, improved models of repeated impressions, and larger pediatric datasets. A powered utility study is also needed before using the synthetic collection as training augmentation; the present dataset is too small to detect a five-point equal-error-rate effect.

## VI. CONCLUSION

This paper evaluated synthetic fingerprints for children under four generated from a model fine-tuned on a small longitudinal pediatric dataset. The proposed protocol combines image quality, minutiae extraction, real-reference similarity, fingerprint-pattern classification, and pairwise synthetic distinctness, followed by a symmetric verification of the retained collection. From 32,000 generated candidates, 1,985 fingerprints remained after final verification. The results also show that a data-derived two-group age representation better matches the younger-group pattern distribution in this dataset, while repeated-impression evaluation reveals differences between minutiae-based and texture-based similarity. Overall, synthetic pediatric fingerprints can provide a useful controlled research resource, but their interpretation should remain tied to the specific generation process, quality criteria, and biometric matchers used in the evaluation.

## ETHICS AND DATA AVAILABILITY

The real fingerprints were collected under an institutional review board-approved protocol with parental consent. No real child's fingerprint is shown in this paper; all displayed examples are synthetic. Because fingerprint recognition in young children has both beneficial and potentially sensitive uses, the synthetic collection, manifest, candidate-evaluation log, and numbers registry are available on request under a data-use agreement and institutional review.